\documentstyle[prb,aps,preprint]{revtex}

\begin{document}

\preprint{DTP 96--63\hskip1cm cond-mat/9612114}
\draft
\title{Transition from $\nu=8/5$ to $\nu=5/3$ in the low
         Zeeman energy limit}

\author{Jacek Dziarmaga\thanks{E-mail: {\tt J.P.Dziarmaga@durham.ac.uk}}
and Meik Hellmund\thanks{E-mail: 
{\tt Meik.Hellmund@durham.ac.uk}}\thanks{
permanent address: Institut
f\"ur Theoretische Physik, Leipzig, Germany}}

\address{Department of Mathematical Sciences,
        University of Durham, South Road, Durham, DH1 3LE,
        United Kingdom}
\date{December 11, 1996}
\maketitle
\tighten

\begin{abstract}
  Skyrmions in the FQHE at filling fractions 
above $\nu=1/3$ are studied within the anyon model and by 
exact diagonalization. Relations to the composite fermion theory are
pointed out. We find that unpolarized quasiparticles above $\nu=1/3$ 
are stable below $B\approx 0.02T$.  At low Zeeman energy the
polarization in the range $\nu=8/5\ldots 5/3$ is found to be a linear 
function of the filling factor. We also reexamine the energy and wave
function of skyrmions at $\nu=1$ by a new method.
\end{abstract}

\section{Introduction}

  The quantum Hall effect continues to surprise. One of the newly found
phenomena are skyrmions\cite{skyrm} around  filling fraction 
$\nu=1$. The question arises naturally if similar 
phenomena could be observed in the fractional quantum Hall regime. 

  In this paper we investigate unpolarized quasiparticles above $\nu=1/3$.
They are constructed within the framework of the anyon model\cite{anyon}.
We found the anyon model to be closely related to the composite fermion
approach\cite{cf}. Our exact diagonalization studies show that 
unpolarized quasiparticles can be stable only below a critical magnetic
field of $0.02T$.  In spite of this they can play a prominent role 
at $\nu$ closer to $2/5$, a state which can be interpreted as their
condensate. In particular
we find that the polarization should be a linear function of the filling
factor in some range below $\nu=2/5$. 

In section~\ref{s2} we reexamine 
  skyrmion wave functions at $\nu=1$  by an exact
diagonalization in  planar geometry. Translational invariance restricts
the number of possible wave functions to just two. One is the wave function
proposed in\cite{sigma} and the other candidate is a uniform state. The
first candidate turns out as the one with lower energy. 
Spin-reversed states around
$\nu=1/3$ are  studied in section~\ref{s3}. It is 
stressed that the unpolarized quasiparticles above $\nu=1/3$ with one
reversed spin are uniform in the thermodynamic limit.  We construct a
wave function basis for the lowest Landau level of such quasiparticles.

In section~\ref{s4} the transition from the spin--singlet $\nu=2/5$ state to
the ferromagnetic state at $\nu=1/3$ is studied in a hard core
model. The polarization appears to be a linear function of the 
filling fraction. We consider the 
  modifications due to a realistic Coulomb potential.
By particle--hole duality this picture  also applies to the transition
from $\nu=8/5$ to $5/3$ which should be more accessible experimentally.

\section{ Skyrmions near $\nu=1$ }
\label{s2}

  Let the operator $a_{l}^{\dagger}\,(b_{l}^{\dagger})$ create a 
spin-up (down)
electron in the $l$-th orbital of the lowest Landau level,
$z^{l}\exp(-|z|^{2}/4)$. The $\nu=1$ ferromagnetic state is
$|1>\equiv\prod_{l=0}^{\infty}b_{l}^{\dagger}\;|0>$. A polarized hole of
angular momentum $L_{z}=h$ can be created by removing one electron from the
orbital $h$, $|h>\equiv b_{h}|1>=\prod_{l\neq h} b_{l}^{\dagger}\;|0>$. 
In the coordinate representation a state with a hole at $w$ is given by

   \begin{equation}\label{j20}
    \Psi_{hole}(w;z_{k})=
    \prod_{k=1}^{\infty}(z_{k}-w)
    \prod_{m>n=1}^{\infty}(z_{m}-z_{n})\;
    |\downarrow_{1}\ldots> \;\;.
   \end{equation}
This is not an eigenstate of the angular momentum $L_{z}$ relative to the
origin. The $|h>$ state of the angular momentum $L_{z}=h$ is obtained from
(\ref{j20}) by the projection

   \begin{equation}\label{j30}
    \Psi_{h}(z_{k}) =
    \int d^{2}w\;e^{-|w|/4} \bar{w}^{h} \Psi_{hole}(w;z_{1},\ldots,z_{N})
    =
    \lim_{w\to 0}\frac{\partial^{h}}{\partial w^{h}}
    \Psi_{hole}(w;z_{1},\ldots,z_{N}) \;\;,
   \end{equation}
where the equality holds up to a normalization factor. By construction all
the states $ \Psi_{h}$ have the same energy provided the two-body
interaction is translationally invariant. For  Coulomb interaction the
gross hole energy is $\varepsilon_{-}=1.25331\frac{e^{2}}{\kappa l}$. 

   The polarized one hole state in the $\nu=1$ ferromagnet is completely
determined by its angular momentum. For low Zeeman energy a number $R$
of electrons  can easily be excited to the up-spin LLL band. Instead of
one polarized state we have now a whole subspace of unpolarized states

   \begin{equation}\label{j40}
   |h_{1},\ldots,h_{R+1};s_{1},\ldots,s_{R}>\equiv
   a^{\dagger}_{s_{1}}\ldots a^{\dagger}_{s_{2}}
   b_{h_{1}}\ldots b_{h_{R+1}}|1>  \;\;,
   \end{equation} 
where $h_{1}<\ldots<h_{R+1}$, $s_{1}<\ldots<s_{R}$ and
$h_{1}+\ldots+h_{R+1}-s_{1}-\ldots-s_{R}=L_{z}=h$. 
The total charge with respect
to the $\nu=1$ state is still $1e$ and the angular momentum $L_{z}$ is the
same as before. Each of the states (\ref{j40}) has a higher energy than the
polarized hole but an appropriate combination of them may  have
less Coulomb energy. 
 
We now restrict ourselves 
  to the case of one reversed spin, $R=1$. The general
form of the textured hole, restricted just by translational invariance,
is

   \begin{equation}\label{j50}
   (z_{1}-w)^{K}
   \prod_{k=2}^{\infty}(z_{k}-z_{1})^{\alpha}(z_{k}-w)^{(2-\alpha)} 
   \prod_{m>n=2}^{\infty}(z_{m}-z_{n})\;
   |\uparrow_{1}\downarrow_{2}\ldots> \;.
   \end{equation}
Note that all the coordinates appear only in  differences of electron
coordinates or in
differences of electron and skyrmion coordinates. There are two holes in
the $\nu=1$ ferromagnet. $(2-\alpha)$ holes are localized at $w$ and
$\alpha$ holes follow the spin-up electron at $z_{1}$. The angular
momentum of the state (\ref{j50}) relative to $w$ is $L=1-K$ .  States of
higher angular momenta with respect to the origin can be generated from
(\ref{j50}) by a projection similar to Eq.(\ref{j30}). A family of
degenerate states can be constructed in this way for each $K\geq 0$. 

   Even for definite angular momentum, there still is some freedom left
in the choice of $\alpha=0,1,2$. For $K=0$ there are three states

   \begin{equation}\label{j60}
   \prod_{k=2}^{N}(z_{k}-z_{1})^{(2-\alpha)} \; (z_{k}-w)^{\alpha} \;
   \prod_{m>n=2}^{N}(z_{m}-z_{n})\;
   |\uparrow_{1}\downarrow_{2}\ldots\downarrow_{N}> \;\;,
   \end{equation}
where we have introduced the regulator $N$. The $\alpha=2$ state is
unstable. In the language of second quantization this is just one state
$|0,1;0>$ of Coulomb energy $1.12\varepsilon_{-}>\varepsilon_{-}$.
Thus we are left with the two cases $\alpha=0$ or $1$. 

  The $\alpha=1$ state has been proposed in\cite{sigma} on the basis that
it is a zero energy eigenstate of the hard-core interaction and that its
spin is $S=\frac{N}{2}-1$ for large $N$. The $\alpha=0$ state is also a
zero energy eigenstate of the hard-core interaction and, as we prove in
Appendix A, it has also  spin $S=\frac{N}{2}-1$. The $\alpha=0$
state can not be restricted to $N$ orbitals since the spin-up electron
coordinate appears with  powers up to $z_{1}^{(2N-2)}$. This is not a
problem for  the field theoretical limit $N=\infty$, where the $\alpha=0$
state is a state with spin reversal. Its charge distribution is 
uniform in accordance with its topological charge equal to zero.

  To verify the  form of the skyrmion wave function we
 performed an exact diagonalization in the space of states
$|h_{1},h_{2};s>$ with $h_{1}+h_{2}-s=1$. The dimension of this
space is infinite in planar geometry. 
To make the Hilbert space
finite we had to impose a cut-off $ h_{1},h_{2},s \leq M$. The system is
thus effectively forced to be ferromagnetic outside the ring of the $M-th$
orbital. The second quantization form of the $\alpha=1$ state (\ref{j50})
is

  \begin{equation}\label{j70}
  \sum_{a=0}^{M-1} \frac{(-1)^{a}}{\sqrt{a+1}} \; |0,a+1;a>  \;\;.
  \end{equation}
We performed the exact diagonalization for a cut-off in the range
$M=1,\ldots,30$. For $M=30$ the overlap of the state (\ref{j60}) with the
ground state is $0.915$. This derivation is mainly due to the suppression
of  the tail of  (\ref{j70}) 
by the ferromagnetic cut-off. The states $|0,a+1;a>$ contribute $0.986$ to
the squared norm of the ground state. The energy is convergent in an
algebraic way, as could have been expected for a state like (\ref{j70}),
which is localized in an algebraic way itself. The extrapolation with a
rational function of $1/M$ to the limit $1/M=0$ gives an $R=1$ skyrmion
energy of $0.9565(1)\;\varepsilon_{-}$. The number in brackets is the
extrapolation error of the last digit. Therefore we expect the $R=1$
skyrmion to become more
stable than the polarized hole at a critical magnetic field of around
$70T$. 

  The imposed cut-off breaks the translational symmetry. In a
translationally invariant system the $\alpha=1$ state (\ref{j50}) is a
seed state for a multiplet of degenerate states, which are generated by
the operator $L^{+}=(\partial_{z_{1}}+\ldots+\partial_{z_{N}})$ or
equivalently by a projection applied to the $\alpha=1$ state
(\ref{j50}) like in Eq.(\ref{j30}). 
We  checked that this symmetry is restored in the
limit of infinite $M$ and that the extrapolation to this limit based on the
data for $M\leq 30$ is justified by  performing similar
diagonalizations in the subspaces of the states $|h_{1},h_{2};s>$ with
$h_{1}+h_{2}-s=L$, for $L=2,\ldots,10$. The energies extrapolated to
$1/M=0$ are listed in table~\ref{T1}.  All the states are degenerate up to
the extrapolation error. This shows that out extrapolation method is
valid in the limit of large $M$ where translational symmetry is restored.
This confirms the assumed form (\ref{j50}) of the wave function. The spin
texture is localized in a power law way. 

  The $\alpha=0$ state of (\ref{j50}) proved to be unstable. This state
has interesting correlations characteristic for the $(1,1,2)$
Laughlin-Halperin state. One could speculate if a condensate of $\alpha=0$
holes in the form of the $(1,1,2)$ Halperin-like state could  be the
ground state at $\nu=2/3$. Unfortunately this state is known to be
unstable in the thermodynamic limit.

\section{ Skyrmions near $\nu=1/3$}
\label{s3}

 Skyrmions near the filling fraction $\nu=1$ can be regarded as
bound states of spin-up electrons with holes in the $\nu=1$ spin-down
ferromagnet. Similarly skyrmions near the ferromagnetic filling fractions
$\nu=1/3,1/5$ can be interpreted as bound states of spin-up electrons with
an appropriate number of anyons. Whereas the former interpretation is exact,
the latter, especially if the anyons are assumed to be pointlike, is an
approximation designed to capture the most relevant degrees of freedom.

  We  studied in detail the simplest case of the quasiparticle
skyrmion above $\nu=1/3$ with one reversed spin $R=1$. The polarized
quasiparticle is a quasielectron of electric charge $-e/3$.
Its gross energy is\cite{mh}
$\varepsilon_{+}^{(1/3)}\approx -0.128 \frac{e^{2}}{\kappa l}$.
 The skyrmion is a bound state of a spin-up electron with two
quasiholes of electric charge $e/3$. The gross energy of the spin-up
electron is zero and the gross energy of the quasihole is\cite{mh}
$\varepsilon_{-}^{(1/3)}= 0.231 \frac{e^{2}}{\kappa l} $. 
The difference
between the energies of the skyrmion and the polarized quasiparticle is
$\Delta\varepsilon=2\varepsilon_{-}^{(1/3)}
                  -\varepsilon_{+}^{(1/3)}
                  +\varepsilon_{int}=0.59 +\varepsilon_{int}$, where
$\varepsilon_{int}$ is the interaction energy of the constituents. The
point-like anyon model is completely defined if we specify the
interaction

  \begin{equation}\label{f20}
  V(z,w_{1},w_{2})=
  \frac{\frac{e^{2}}{9\kappa}}{|w_{1}-w_{2}|}-
  \frac{\frac{e^{2}}{3\kappa}}{|z-w_{1}|}-
  \frac{\frac{e^{2}}{3\kappa}}{|z-w_{1}|} \;\;,
  \end{equation}
where $z$ is the coordinate of the spin-up electron and $w_i$ are the
quasihole coordinates. If the electron and the anyons are restricted to
the lowest Landau level, their Hilbert space can be spanned by the
orthonormal basis

  \begin{equation}\label{f30}
  e_{abc}(z,w_{1},w_{2})=
  N_{abc}
  z^{a} e^{-\frac{ |z|^{2} }{ 4 }}
  (\bar{w}_{1}-\bar{w}_{2})^{\frac{1}{3}+2b}
  (\bar{w}_{1}+\bar{w}_{2})^{c}
  e^{ -\frac{ |w_{1}|^{2} }{ 12 } - \frac{ |w_{2}|^{2} }{ 12 } } \;\;.
  \end{equation}
The normalization factors $N_{abc}$ and the matrix elements of the Coulomb
interaction (\ref{f20}) in the basis (\ref{f30}) can be found in Appendix
B. The state $e_{000}$ has the lowest energy in the basis (\ref{f30}). Its
angular momentum is $L_{z}=1/3$. 
To understand how this state mixes with other
states in order to form a skyrmion state we performed an exact
diagonalization in the subspace $L_{z}\equiv 1/3+2b+c-a=1/3$ 
of (\ref{f30}). For this subspace to be finite, we
had to impose the cut-off $a,2b,c\leq M$. The maximal cut-off was $M=17$.
We found that $0.9999$ of the norm squared of the ground state lies
in the subspace spanned by the orthonormal basis

  \begin{equation}\label{f40}
  e_{k}^{(0)}(z,w_{1},w_{2})=
  N_{k}^{(0)}
  z^{k} e^{-\frac{ |z|^{2} }{ 4 }}
  (\bar{w}_{1}-\bar{w}_{2})^{\frac{1}{3}}
  (\bar{w}_{1}+\bar{w}_{2})^{k}
  e^{ -\frac{ |w_{1}|^{2} }{ 12 } - \frac{ |w_{2}|^{2} }{ 12 } } \;\;.
  \end{equation}
The two anyons are found to be in the state of lowest possible
relative angular momentum  $1/3$. It seems as if there were just two
relevant particles in the problem: the spin-up electron at $z$ and a
double quasihole at $w_{1}+w_{2}$. The attraction between the electron and
the anyons dominates the mutual repulsion between anyons (\ref{f20}).
The two anyons try to be as close as possible to the spin-up
electron even if this costs some (small) amount of repulsion energy. 
The spin-up
electron coordinate must appear in the ground state wave function as
$\prod_{k=2}^{\infty}
    (z_{k}-z_{1})^{2}
    \prod_{m>n=2}^{\infty}(z_{m}-z_{n})^{3}
    |\uparrow_{1}\downarrow_{2}\ldots>$:  for the spin-down electrons
the spin-up electron appears as a particle with two attached flux quanta. 

   Before we proceed, let us try to determine possible skyrmion wave
functions on general grounds. Similarly as around $\nu=1$ the
translational invariance restricts the wave function to
     
  \begin{equation}\label{f10}
  (z_{1}-w)^{K}
  \prod_{k=2}^{\infty}
  (z_{k}-z_{1})^{(2-\alpha)}
  (z_{k}-w)^{\alpha}
  \prod_{m>n=2}^{\infty}(z_{m}-z_{n})^{3}
  |\uparrow_{1}\downarrow_{2}\ldots>  \;\;.
  \end{equation}
There are two Laughlin quasiholes in addition to the electron with
reversed spin. $\alpha$ quasiholes are localized around $w$ and
$(2-\alpha)$ quasiholes follow the spin-up electron at $z_{1}$. The
angular momentum relative to $w$ is $L_{z}=\frac{1}{3}-K$ as compared to
the $\nu=1/3$ state. The ground state which we found within the anyon model
for $L_{z}=1/3$ can be identified with the $K=0$ and $\alpha=0$ state

 \begin{eqnarray}\label{f15}
 && \prod_{k=2}^{\infty}
    (z_{k}-z_{1})^{2}
    \prod_{m>n=2}^{\infty}(z_{m}-z_{n})^{3}
    |\uparrow_{1}\downarrow_{2}\ldots> =\nonumber\\ &&= 
   [\prod_{i>j=2}^{\infty}(z_{m}-z_{n})]^{2}
     \{ \prod_{k=2}^{\infty}
        (z_{k}-z_{1})^{2}   
        \prod_{m>n=2}^{\infty}(z_{m}-z_{n})^{3}
        |\uparrow_{1}\downarrow_{2}\ldots> \}  \;\;.
  \end{eqnarray}
The equation shows  the way this state is constructed in the composite
fermion prescription from the seed state with filled spin-down LLL and one
electron in the spin-up LLL. The spin and charge distributions of this
state are uniform. We discuss quasiparticle properties of the state
(\ref{f15}) in section \ref{ss31}. 

  If we restrict ourselves to $K=0$ it is easy to understand 
why $\alpha=0$ is preferred on energetic grounds. The $\alpha=2$ 
state is just the single
$e_{000}$ state of the anyon model. This state is not an eigenstate of the
Coulomb interaction (see Appendix B), so it can not be the ground state. For
$\alpha=1,w=0$ the wave function (\ref{f10}) can be expanded in powers of
$z_{1}$

  \begin{eqnarray}\label{f50}
  &&\prod_{k=2}^{\infty}
    (z_{k}-z_{1}) 
    z_{k}
    \prod_{m>n=2}^{\infty}(z_{m}-z_{n})^{3}
    |\uparrow_{1}\downarrow_{2}\ldots>=   \nonumber \\
  &&[ \prod_{k=2}^{\infty} z_{k}^{2}-
      z_{1}\sum_{l=2}^{\infty} \prod_{\stackrel{k=2,3,4,\ldots}
                                               {k\neq l}} z_{k}^{2}+\ldots]
    \prod_{m>n=2}^{\infty}(z_{m}-z_{n})^{3}
    |\uparrow_{1}\downarrow_{2}\ldots>  \;\;.
  \end{eqnarray}
The subsequent terms in this expansion are proportional to the following
states in the anyon model

  \begin{equation}\label{f60}
  \tilde{e}_{k}(z,w_{1},w_{2})=
  \tilde{N}_{k}
  z^{k} e^{-\frac{ |z|^{2} }{ 4 }}
  (\bar{w}_{1}-\bar{w}_{2})^{\frac{1}{3}}
  (\bar{w}_{1}^{k}+\bar{w}_{2}^{k})
  e^{ -\frac{ |w_{1}|^{2} }{ 12 } - \frac{ |w_{2}|^{2} }{ 12 } } \;\;.
  \end{equation}
A similar expansion in powers of $z_{1}$ for the $\alpha=1,w=0$ wave
function (\ref{f10}) gives rise to the states (\ref{f40}).  In the $k$-th
state of the family (\ref{f60}) the spin-up electron sits in the $k$-th
electronic orbital, one of the anyons is in the $0$-th anyonic orbital and
the other anyon sits in the $k$-th anyonic orbital. As the hole's magnetic
length is $\sqrt{3}$ that of the electron, the $k$-th electronic and
anyonic orbitals soon become quite distant with increasing $k$.  The
attraction between the electron and each of the anyons becomes weaker 
 with increasing $k$. On the other hand in the $k$-th state of the
family (\ref{f40}) there is the spin-up electron in the $k$-th orbital and
the double hole in the $k$-th double hole orbital. The double hole's
magnetic length is just $\sqrt{3/2}$ that of the electron, so that the
$k$-th orbitals are now much closer to each other. The attraction decreases
with increasing $k$ but much slower than for the states (\ref{f60}).  The
general rule is that for an electron interacting with $n$ anyons of
electric charge $e/p$, the anyons will tend to group into a $p$-particle
cluster and $(n-p)$ single anyons. To illustrate this rule let us consider
the quasihole skyrmion just below $\nu=1/3$. Its general form 
which is restricted just by translational invariance is

  \begin{equation}\label{f70}
  \prod_{k=2}^{\infty}
  (z_{k}-z_{1})^{(4-\alpha)}
  (z_{k}-w)^{\alpha}
  \prod_{m>n=2}^{\infty}(z_{m}-z_{n})^{3}
  |\uparrow_{1}\downarrow_{2}\ldots>  \;\;.
  \end{equation}
There are one spin-up electron and four quasiholes. The rule predicts the
$\alpha=1$ state to be the ground state as it is indeed the case\cite{jk}. 
The rule also gives the correct wave functions for skyrmions around
$\nu=1/5$. 

  In the special case of $p=1$, the rule predicts that the $\alpha=1$
state (\ref{j60}) is more stable than the $\alpha=0$ state (\ref{j60}).

\subsection{ Lowest Landau level of the unpolarized quasiparticle }
\label{ss31}

  The state (\ref{f15}) is just one of the family of degenerate states

  \begin{eqnarray}\label{x50}
  &&\Psi_{K}(w,z_{k})=
    (z_{1}-w)^{K}
    \prod_{k=2}^{\infty}
    (z_{k}-z_{1})^{2}
    \prod_{m>n=2}^{\infty}(z_{m}-z_{n})^{3} \;
    |\uparrow_{1}\downarrow_{2}\ldots>=   \nonumber \\
  &&[\prod_{i>j=1}^{\infty}(z_{i}-z_{j})]^{2}\;
    \{\;(z_{1}-w)^{K}
        \prod_{k=2}^{\infty}
        (z_{k}-z_{1})^{2}
        \prod_{m>n=2}^{\infty}(z_{m}-z_{n}) \;
        |\uparrow_{1}\downarrow_{2}\ldots> \;\}  \;\;,
  \end{eqnarray}
with $K=0,1,2\ldots$. The states are degenerate because they are related by 
 
  \begin{equation}\label{x60}
  \Psi_{K-1}(w,z_{k})=-\frac{1}{K}\partial_{w} \Psi_{K}(w,z_{k})=
                      \frac{1}{K} L^{+}  \Psi_{K}(w,z_{k}) \;\;
  \end{equation}
and the energy should not depend on the choice of $w$.
The anyon model, which is by construction in the field
theoretical limit $N=\infty$, has this particular symmetry too. 
We checked this by exact diagonalizations in the subspaces spanned by the
orthonormal states

  \begin{equation}\label{x70}
  e_{k}^{(K)}(z,w_{1},w_{2})=
  N_{k}^{(K)}
  z^{k+K} e^{-\frac{ |z|^{2} }{ 4 }}
  (\bar{w}_{1}-\bar{w}_{2})^{\frac{1}{3}}
  (\bar{w}_{1}+\bar{w}_{2})^{k}
  e^{ -\frac{ |w_{1}|^{2} }{ 12 } - \frac{ |w_{2}|^{2} }{ 12 } } \;\;
  \end{equation}
in the range of $K=0,\ldots,6$. The interaction energies extrapolated to
$1/M=0$ are listed in table~\ref{t2}. They are degenerate up to extrapolation
errors. Thus the anyon model correctly reproduces the global symmetry. 

  In the state $\Psi_{K}$ the spin-up electron has been removed from the
orbitals $0,\ldots,K-1$. The states (\ref{x50}) can be interpreted as 
orbitals in the lowest Landau level of the unpolarized quasielectron. They
are obtained in the CF prescription from a filled spin-down LLL and a
spin-up electron in the $K$-th orbital of the spin-up LLL.

\subsection{ Energy of the unpolarized quasiparticle }
\label{ss32}

To get an estimate for the energy gain by depolarization we
made a finite size study of electrons in spherical geometry\cite{Hald83}.
The one quasiparticle sector at $\nu=1/3$ is characterized by $m=3(N-1)-1$
flux quanta piercing the sphere. 
We calculated the lowest eigenstates for Coulomb interaction in the 
subspace of states with $S_z=N/2-1$ for up to 9 electrons.  
The lowest energy state has always $S=N/2-1$, the next state $S=N/2$.
The energies are given in table~\ref{t3}. The data allow fits with 
quadratic polynomials in $1/N$ which gives a gap energy for $N\to\infty$ 
of $0.000\,9(3) e^2/\varepsilon l$.
 
Therefore, the unpolarized quasielectron is found to be stable but only at
marginally small magnetic fields $B<0.02T$. It is even less likely to be
observed if we take impurities into account. The polarized quasielectron
is a localized object, which can form a bound state with an impurity. The
unpolarized quasielectron (\ref{f15}) is uniform so its interaction energy
with a localized potential is zero to first approximation. The stability
of a fluid of quasielectrons can be enhanced when they condense into a
Laughlin fluid. In fact, the stable unpolarized $\nu=2/5$ state can be
interpreted as such a condensate.

\section{ Transition from $\nu=8/5$ to $\nu=5/3$  }
\label{s4}

   The ferromagnetic $\nu=1/3$ ground state is described by the celebrated
Laughlin wave function

   \begin{equation}\label{10}
   \psi_{1/3}(z_{k})=
   \prod_{m>n=1}^{N}(z_{m}-z_{n})^{3} \; 
   |\downarrow_{1}\ldots\downarrow_{N}>=
   [\prod_{k>l=1}^{N}(z_{k}-z_{l})]^{2} \;
   \{ \; \prod_{m>n=1}^{N}(z_{m}-z_{n}) \; 
      |\downarrow_{1}\ldots\downarrow_{N}> \} \;\;.  
   \end{equation}
The second equality shows how this state is constructed from the $\nu=1$
ferromagnet in the composite fermion theory. In the same framework, the
spin-singlet $\nu=2/5$ is obtained from the spin-singlet $\nu=2$ state,

   \begin{eqnarray}\label{20}
   \psi_{2/5}(z_{k})&=&
   \prod_{i>j=1}^{N/2}(z_{i}-z_{j})^{3}
   \prod_{k>l=N/2}^{N}(z_{k}-z_{l})^{3}
   \prod_{m=0}^{N/2} \prod_{n=N/2}^{N} (z_{m}-z_{n})^{2} \;
   |\uparrow_{1}\ldots\uparrow_{N/2}\downarrow_{N/2+1}\ldots\downarrow_{N}>=
                                                           \nonumber \\
   &&[\prod_{i>j=1}^{N}(z_{i}-z_{j})]^{2} \;\;
     \{ \; \prod_{k>l=1}^{N/2}(z_{k}-z_{l})
           \prod_{m>n=N/2}^{N}(z_{m}-z_{n}) \; 
   |\uparrow_{1}\ldots\uparrow_{N/2}\downarrow_{N/2+1}\ldots\downarrow_{N}> \} 
   \;\;,
   \end{eqnarray}
where we have chosen the spinor component with the spins $1,..,N/2$ up and
the spins $N/2+1,\ldots,N$ down. The spin-singlet $\nu=2/5$ state is the
$(3,3,2)$ Halperin wave function. The states (\ref{10}) and (\ref{20}) are
both zero energy eigenstates of the hard core model 

  \begin{equation}\label{25}
  V(z)=\infty \delta(z)+\lambda\nabla^{2}\delta(z) 
  \end{equation}
with pseudopotentials $V_{0}=\infty,V_{1}>0$ and $V_{k}=0$ for $k\geq 2$. 

  A fully polarized state at $\nu >1/3$ can not have zero energy because the
electrons are packed too close. More space can be created by
reversing a number of spins. An example of a zero energy eigenstate one
flux quantum above $\nu=1/3$ is the unpolarized quasiparticle (\ref{f15})
with one reversed spin

  \begin{equation}\label{30}
  \prod_{k=2}^{N} 
  (z_{k}-z_{1})^{2}
  \prod_{m>n=2}^{N} (z_{m}-z_{n})^{3} \;
  |\uparrow_{1}\downarrow_{2}\ldots\downarrow_{N}> \;\;.
  \end{equation}
The spin of the state (\ref{30}) is $S=\frac{N}{2}-1$. 

  Two flux quanta above $1/3$ the polarized ground state would contain two
quasielectrons. To have zero energy, a number of spins must flip. For
just one flipped spin a candidate for a skyrmion is

  \begin{equation}\label{40}
  \prod_{k=2}^{N}
  (z_{k}-z_{1})
  \prod_{m>n=2}^{N} (z_{m}-z_{n})^{3} \;
  |\uparrow_{1}\downarrow_{2}\ldots\downarrow_{N}> \;\;.
  \end{equation}
The energy of this state does not vanish. The zero energy state must
contain at least $R=2$ reversed spins, for example

  \begin{equation}\label{50}
  \prod_{i>j=1}^{2} (z_{i}-z_{j})^{3}
  \prod_{k=3}^{N}
  \prod_{l=1}^{2}
  (z_{k}-z_{l})^{2}
  \prod_{m>n=3}^{N} (z_{m}-z_{n})^{3} \;
  |\uparrow_{1}\uparrow_{2}\downarrow_{3}\ldots\downarrow_{N}> \;\;.
  \end{equation} 
The two examples above can be  easily generalized. To construct a zero
energy eigenstate $\Phi$ flux quanta above the filling fraction $1/3$ one
has to flip at least $\Phi$ spins, $R\geq\Phi$. An example of zero energy
state is

  \begin{equation}\label{55}
  \prod_{i>j=1}^{\Phi} (z_{i}-z_{j})^{3}
  \prod_{k=\Phi+1}^{N}
  \prod_{l=1}^{\Phi}   
  (z_{k}-z_{l})^{2}
  \prod_{m>n=\Phi+1}^{N} (z_{m}-z_{n})^{3} \;
  |\uparrow_{1}\ldots\uparrow_{\Phi}\downarrow_{\Phi+1}\ldots
      \downarrow_{N}> \;\;.
  \end{equation}
The spin of the above state is $S=\frac{N}{2}-\Phi$. The zero energy
states have spins in the range of $S=\frac{N}{2}-\Phi,\ldots,0$. A state with
$S<\frac{N}{2}-\Phi$ can not have zero energy. In particular the
spin-singlet state (\ref{20}) is a unique zero energy state at
$\nu=2/5$, where $\Phi=\frac{N}{2}$ . 

  So far the Zeeman energy was set to zero. The ground states for
$S<\frac{N}{2}-\Phi$ have positive potential energies proportional to the
only relevant quasipotential $V_{1}$, let us call them $V_{1}e(S)$. The
total energies of the ground states are $E(S)=V_{1}e(S)-GS$, where $G$ is
the Zeeman energy. The ground state is polarized $(S=\frac{N}{2})$ for
sufficiently large $G$. As the Zeeman energy is decreased, the
polarization $P\equiv 2S/N$ decreases from $P=1$ to $P\leq
\frac{(2/5-\nu)}{(2/5-1/3)}$. It can not be any less because
$e(0)=\ldots=e(\frac{N}{2}-\Phi)$ so that for any $G>0$ we have
$E(\frac{N}{2}-\Phi)<\ldots<E(0)$. 

  We conclude that in the hard core model (\ref{25}) in the limit of
small Zeeman energy $G=0^{+}$ the polarization depends linearly on the
filling fraction
 
  \begin{equation}\label{60}
  P=\frac{(2/5-\nu)}{(2/5-1/3)}
  \end{equation}
in the range $\nu\in (1/3,2/5)$. Similar results hold, by particle-hole
symmetry, for $\nu\in (8/5,5/3)$. 

  How is this result modified by a more realistic interaction potential
and what is its experimental significance? We have seen that the
unpolarized quasiparticle (\ref{30}) is unstable for realistic magnetic
fields. On the other hand, the spin-singlet $\nu=8/5$ state is
experimentally observed\cite{exp} at fields of around $B=5T$. Thus at
$\nu=8/5$ the low Zeeman energy limit can be achieved in present
experiments. The polarization is bounded from below by (\ref{60}) since
the zero energy band is exactly degenerate in the hard core model. This
degeneracy is slightly removed for Coulomb interaction but without
decreasing the polarization below (\ref{60}) at realistic magnetic fields.
Thus we expect that the polarization follows the linear pattern (\ref{60})
in some range above $\nu=8/5$ before it increases faster than linear to
$P=1$. 

  To understand better why the degeneracy of the hard core zero energy
band is only slighly removed by the Coulomb interaction, let us observe
that all the hard core zero energy states below $\nu=2/5$ can be obtained
by inserting quasiholes into the spin-up or spin-down fluids of the
(3,3,2) state. The quasiholes interact by Coulomb potential but their
electric charge is only $e/5$. Let us consider the example of two
quasiholes. The spin can be either $S=0$ or $S=1$. The lowest state for a
given spin is the one with the highest relative angular momentum of the
quasiparticles. For a given number of flux quanta the highest relative
angular momenta for $S=0$ and $S=1$ differ only by $1$. If these maximal
relative angular momenta are large (low quasihole density) the
corresponding states are almost degenerate. Degeneracy is removed by the
Zeeman energy, which favors the $S=1$ ground state. For more than two
quasiholes the ground state for any given spin is the state with the
highest possible pairs' angular momenta. By a similar argument as for two
holes, the polarization below $\nu=2/5$ (or above $\nu=8/5$) is given by
(\ref{60}) for low Zeeman energy.

\acknowledgements

J.D. is supported by UK PPARC and M.H. by Deutscher Akademischer
Austauschdienst.

\appendix

\section{}

  The proof is a generalization of an analogous argument in\cite{sigma}. 
For any wave-function $\psi$ its "bosonic" part $\psi_{B}$ can be defined
by $\psi=\psi_{B}\prod_{m>n=1}^{N}(z_{m}-z_{n})$. The spin quantum numbers
of $\psi$ and $\psi_{B}$ are the same. The bosonic part of the $\alpha=0$
state (\ref{j50}) is

   \begin{equation}\label{a10}
   \prod_{k=2}^{N}(z_{k}-z_{1})\;
   |\uparrow_{1}\downarrow_{2}\ldots\downarrow_{N}>\equiv
   \phi_{1}(\{z_{l}\})\;
   |\uparrow_{1}\downarrow_{2}\ldots\downarrow_{N}> \;\;.
   \end{equation}
As usual, the exponential factors are omitted and only one spinor
component is shown. Other spinor components can be obtained by
symmetrization. The expectation value of the operator

   \begin{equation}\label{a20}
   S^{2}=S_{z}^{2}+\frac{S_{+}S_{-}+S_{-}S_{+}}{2}
   \end{equation}
in state (\ref{a10}) can be worked out as

   \begin{equation}\label{a30}
   <S^{2}>=(\frac{N}{2}-1)^{2}+\frac{N}{2}+
           (N-1)\frac{<\phi_{1}|\phi_{2}>}
                     {||\phi_{1}||\;||\phi_{2}||} \;\;.
   \end{equation}
The scalar product is understood as the integral
$<f,g>=\int\prod_{k=1}^{N}d^{2}z_{k}\exp(-|z_{k}|^{2}/2) f^{\star}g$. 
By symmetry $||\phi_{1}||=||\phi_{2}||$.
The norm can be found to be

   \begin{equation}\label{a40}
   ||\phi_{1}||^{2}=
   \pi^{N}2^{2N-1}(N-1)!\sum_{s=0}^{N-1}\frac{1}{s!}
   \end{equation}
and the unnormalized overlap

   \begin{equation}\label{a50}
   <\phi_{1}|\phi_{2}>=\pi^{N}2^{2N-1}N \;\;,
   \end{equation}
so that

   \begin{equation}\label{a60}
   <S^{2}> \equiv S(S+1) = \nonumber \\
   (\frac{N}{2}-1)^{2}+\frac{N}{2}+
   \frac{N}{ (N-2)!\sum_{s=0}^{N-1}\frac{1}{s!} }\;\; .
   \end{equation}
Equation (\ref{a60}) can be solved with respect to $S$. The result is
$S=\frac{N}{2}-1+O(1/N)$, which was to be demonstrated. The state
(\ref{a10}) has $S_{z}=-\frac{N}{2}+1$. As such it is in general a
combination of eigenstates with $S=\frac{N}{2}-1$ and $\frac{N}{2}$. 
In the thermodynamic limit only the $S=\frac{N}{2}-1$ states survives.

\section{}

The normalization factor of the states (\ref{f30}) is

  \begin{equation}\label{B10}
  N^{-2}_{abc}=  \pi^{3} \;
            2^{(a-1)} \;
            12^{(2b+c+\frac{7}{3})} \;
            \Gamma[a+1] \;
            \Gamma[2b+\frac{4}{3}] \;
            \Gamma[c+1]  \;\;.
  \end{equation}
The matrix elements of the interaction (\ref{f20}) in the basis
(\ref{f30}) are

  \begin{eqnarray}\label{B20}
 && <e_{abc}|V|e_{def}>=
  \frac{\delta_{ad}\;\delta_{be}\;\delta_{cf}\;\Gamma[2b+\frac{5}{6}]}
       {9\;\sqrt{12}\;\Gamma[2b+\frac{4}{3}]}-\nonumber\\&&-
  \frac{1}{6} \; N_{abc} \; N_{def} \; A_{ad} \; B_{be} \; C_{cf} \;
  I[a+d,b+e,c+f,|a-d|,|b-e|,|c-f|] \;\;,
  \end{eqnarray}
where the various factors are

  \begin{eqnarray}\label{B30}
  &&A_{ad}=\pi \; i^{|a-d|} \; 2^{(\frac{a+d}{2}-\frac{|a-d|}{2}+1)} \;
           \frac{\Gamma[\frac{a+d}{2}+\frac{|a-d|}{2}+1]}
                {\Gamma[|a-d|+1]}                      \;\;,\nonumber\\
  &&B_{be}=\pi \; (-i)^{|b-e|} \; 12^{(\frac{b+e}{2}+1)} \;
           (\frac{3}{4})^{\frac{|b-e|}{2}} \;
            \frac{\Gamma[\frac{b+e}{2}+\frac{|b-e|}{2}+1]}
                 {\Gamma[|b-e|+1]}                      \;\;,\nonumber\\
  &&C_{cf}=\pi \; i^{2|c-f|} \; 12^{(c+f+\frac{4}{3})} \;
           (\frac{3}{4})^{|c-f|} \;
           \frac{\Gamma[a+d+|a-d|+\frac{4}{3}]}
                {\Gamma[2|c-f|+1]}                      \;\;,\nonumber\\
  && I[A,B,C,a,b,c]=                                         \nonumber\\
  && \int_{0}^{\infty} dq \; \frac{q^{(a+2b+c)}}{\exp(2q^{2})} \;
                    _{1}F_{1}[\frac{a-A}{2},a+1,\frac{q^{2}}{2}] \;
                    _{1}F_{1}[\frac{b-B}{2},b+1,\frac{3q^{2}}{2}] \;
                    _{1}F_{1}[c-C-\frac{1}{3},2a+1,\frac{3q^{2}}{2}] \;\;.
   \end{eqnarray}

\section{}

The normalization factors of the basis (\ref{x70}) are

   \begin{equation}\label{c10}
   \left(N_{k}^{(K)}\right)^{-2}= \pi^{3} \;
                 2^{(K+k-1)} \;
                 12^{(k+\frac{7}{3})} \;
                 \Gamma[ \frac{4}{3} ] \;
                 \Gamma[ k+1] \;
                 \Gamma[ K+k+1 ]   \;\;.
    \end{equation}
The matrix elements of the interaction (\ref{f20}) in the basis
(\ref{x70}) are

   \begin{eqnarray}\label{c20}
   && <e_{k}^{(K)}|V|e_{l}^{(K)}>=
                \frac{\Gamma[\frac{5}{6}]}
                     {9 \; \sqrt{12} \; \Gamma[\frac{4}{3}] } 
    -2\pi^{3} \; N_{k}^{(K)} \; N_{l}^{(K)} J_{K}[k+l,|k-l|] 
       \nonumber\\ &&\times
      \frac{ 2^{(\frac{k+l-|k-l|}{2}+K+1)} \;
             12^{(\frac{k+l}{2}+\frac{4}{3})} \;
             (\frac{4}{3})^{\frac{|k-l|}{2}} \;
             \Gamma[\frac{4}{3}] \;
             \Gamma[\frac{k+l+|k-l|}{2}+1] \;
             \Gamma[\frac{k+l+|k-l|}{2}+1+K] }
           { \Gamma^{2}[|k-l|+1] } ,
   \end{eqnarray}
where 

   \begin{equation}\label{c30}
   J_{K}[A,a]=\int_{0}^{\infty} dq \; \frac{ q^{2a} }{ \exp(2q^{2}) } \;
              _{1}F_{1}[-\frac{1}{3},1,\frac{3q^{2}}{2}] \;
              _{1}F_{1}[\frac{a-A}{2},a+1,\frac{3q^{2}}{2}] \;
              _{1}F_{1}[\frac{a-A}{2}-K,a+1,\frac{q^{2}}{2}] \;\;.
   \end{equation}

\begin{table}
  \caption{$R=1, \nu=1$ Skyrmion energy for different angular momenta}
  \label{T1}
\begin{tabular}{cc}
     L          &     $E(L)$ $[\varepsilon_{-}]$   \\ \hline
     1          &     0.9565(1)   \\ 
     2          &     0.9564(2)   \\ 
     3          &     0.9564(1)   \\ 
     4          &     0.9564(1)   \\ 
     5          &     0.9564(1)   \\ 
     6          &     0.9565(1)   \\ 
     7          &     0.9565(1)   \\ 
     8          &     0.9567(3)   \\ 
     9          &     0.9564(2)   \\ 
     10         &     0.9566(3)  
\end{tabular}
\end{table}

\begin{table}
  \caption{Energy of the unpolarized anyonic quasiparticle at $R=1,\nu=1/3$}
  \label{t2}
\begin{tabular}{cc}
     K          &   $E_{int}(K)$ $[\frac{e^{2}}{\kappa l}]$  \\ \hline
     0          &     -0.395(1)   \\ 
     1          &     -0.394(1)   \\ 
     2          &     -0.395(2)   \\ 
     3          &     -0.392(2)   \\ 
     4          &     -0.392(2)   \\ 
     5          &     -0.395(1)   \\ 
     6          &     -0.393(1)  
\end{tabular}
\end{table}

\begin{table}
  \caption{Energy per particle of the lowest states of $N$ electrons
 on a sphere with spin $S=N/2$ and $N/2-1$ at filling fraction 
 ``$1/3+1\;\text{quasiparicle}$''. 
  The energy is in units of $e^2/\varepsilon l'$ with 
$l'$ including a finite size correction\protect\cite{DM89}
  $l'=l\sqrt{\frac{m \nu}{N}}$.}
  \label{t3}
\begin{tabular}{ccc}
N    &    E (S=N/2-1)   &E (S= N/2)      \\ \hline
5    &-0.405418 & -0.397353\\
6    &-0.406033 & -0.399790\\
7    &-0.406445 & -0.401230\\
8    &-0.406795 & -0.402361\\
9    &-0.407090 & -0.403246\\
$\infty$  &-0.4095(1)&-0.4086(2) 
\end{tabular}
\end{table}


\begin{thebibliography}{10}
\bibitem{skyrm} S.~L.~Sondhi, A.~Karlshede, S.~A.~Kivelson and E.~H.~Rezayi,
                Phys. Rev. B {\bf 47}, 16419 (1995);
                S.~E.~Barrett, G.~Dabbagh, L.~N.~Pfeiferer, K.~W.~West
                and R.~Tycko, Phys. Rev. Lett. {\bf 74}, 5112 (1995);
                R.~Tycko, S.~E.~Barrett, G.~Dabbagh, L.~N.~Pfeiferer and
                K.~W.~West, Science {\bf 268}, 1460 (1995).
\bibitem{anyon} F.~D.~M. Haldane, Phys. Rev. Lett. {\bf 51}, 605 (1983);
                B.~I.~Halperin, Phys. Rev. Lett. {\bf 52}, 1583 (1984);
                for a recent application see:
                M.~E.~Portnoi and E.~I.~Rashba, preprint "Theory
                of anyon excitons: Relation to excitons of
                $\nu=1/3$ and $\nu=2/3$ incompressible liquids",
                cond-mat/9608125 (to be published in Phys. Rev. B).
\bibitem{cf}    J.~K.~Jain, Phys. Rev. Lett. {\bf 63}, 199 (1989),
                Adv. Phys. {\bf 41}, 105 (1992).
\bibitem{sigma} A.~H.~MacDonald, H.~A.~Fertig and L.~Brey, 
                Phys. Rev. Lett. {\bf 76}, 2153 (1996).
\bibitem{mh}    R.~Morf and B.~I.~Halperin, Phys. Rev. B {\bf 33}, 2221 (1986).
\bibitem{jk}    R.~K.~Kamilla, X.~G.~Wu and J.~K.~Jain, 
                Solid State Commun. {\bf 99}, 289 (1996).
\bibitem{Hald83}F.~D.~M. Haldane, Phys. Rev. Lett. {\bf 51},  605  (1983).
\bibitem{exp}   J.~P.~Eisenstein, H.~L.~Stormer, L.~N.~Pfeiffer and 
                K.~W.~West, Phys. Rev. Lett. {\bf 62}, 1540 (1989).
\bibitem{DM89}  N.~d'Ambrumenil and R.~Morf, 
                Phys. Rev. B {\bf 40},  6108  (1989).


\end{thebibliography}
\end{document}